\documentclass[12pt]{article}
\usepackage{amssymb,amsmath,epsfig}
\usepackage{graphicx}
\allowdisplaybreaks
\begin{document}

\title{\bf Complexity of Dynamical Sphere in Self-interacting Brans-Dicke Gravity}

\author{M. Sharif \thanks{msharif.math@pu.edu.pk} and Amal Majid
\thanks{amalmajid89@gmail.com}\\
Department of Mathematics, University of the Punjab,\\
Quaid-e-Azam Campus, Lahore-54590, Pakistan.}

\date{}
\maketitle
\begin{abstract}
This paper aims to derive a definition of complexity for a dynamic
spherical system in the background of self-interacting Brans-Dicke
gravity. We measure complexity of the structure in terms of
inhomogeneous energy density, anisotropic pressure and massive
scalar field. For this purpose, we formulate structure scalars by
orthogonally splitting the Riemann tensor. We show that
self-gravitating models collapsing homologously follow the simplest
mode of evolution. Furthermore, we demonstrate the effect of scalar
field on the complexity and evolution of non-dissipative as well as
dissipative systems. The criteria under which the system deviates
from the initial state of zero complexity is also discussed. It is
concluded that complexity of the sphere increases in
self-interacting Brans-Dicke gravity because the homologous model is
not shear-free.
\end{abstract}

\section{Introduction}

Numerous astrophysical experiments (Sloan Digital Sky Survey, Large
Synoptic Survey Telescope, Two-degree Field Galaxy Redshift Survey)
have proved that the mechanism and evolution of the vast universe
massively depend on the large scale structures such as stars and
galaxies. Thus, the study of these components is vital to gain a
better understanding of the cosmos and its origin. These
self-gravitating bodies are intricate cosmic objects whose physical
properties may undergo a fundamental change due to a slight
fluctuation in the interior. Thus, it is necessary to accommodate
all the factors contributing to their complicated nature in a
relation termed as complexity factor. Such a factor depicts the
inter-relationship of various physical parameters (density,
pressure, heat dissipation, etc.) as well as gauges the effects of
external or internal perturbations on the matter variables. It can
also be used to develop a criterion of stability to compare the
complexity of different self-gravitating structures. Researchers
have explored the concept of complexity many times but a standard
definition has not been established \cite{1}.

Previous definitions of complexity were proposed based on
information (distances and symmetries of atomic arrangement) and
entropy (quantification of geometrical attributes) of celestial
systems. However, these definitions failed to accurately evaluate
the complexity of two simple models: ideal gas and perfect crystal.
Since the atoms in ideal gas do not occupy fixed positions,
therefore, maximum information is required to completely specify any
of its probable states. On the other hand, a perfect crystal
requires minimum information for its description due to the
symmetrical arrangement of atoms. Despite the differences in their
geometrical configurations, both physical systems demonstrate
minimum complexity. Lopez-Ruiz et al. \cite{5} improved the earlier
definitions by encompassing the concept of disequilibrium. The main
idea was to detect how various probable states differ from the
equiprobable distribution of the physical structure. According to
this definition, the two systems (ideal gas and a perfect crystal)
have zero complexity. The complexity of neutron stars and white
dwarfs has been computed by using energy density in place of
probability distribution in the above definition \cite{9}.

Dense stellar systems have tightly packed particles in their
interior. This arrangement restricts the movement of nuclear matter
in the radial direction. Consequently, radial pressure is less than
the force in the transverse direction leading to anisotropy in
pressure. Thus, anisotropy plays a significant role in determining
the viability and stability of self-gravitating systems. Since the
definition proposed by Lopez-Ruiz et al. incorporates energy density
only and lacks the contribution of other state determinants (such as
anisotropy) therefore, it cannot be considered as the best criterion
of complexity. Recently, Herrera \cite{13} devised a new complexity
factor for static sphere in the context of general relativity (GR)
by assuming that the complexity-free system is isotropic as well as
homogeneous. The distinguishing feature of Herrera's technique is
the integration of the system's active gravitational mass,
inhomogeneous energy density and anisotropic pressure in the
definition of complexity. He obtained structure scalars through the
orthogonal splitting of Riemann tensor to generate the complexity
factor.

Herrera's definition of complexity has also been extended for a
non-static radiating sphere by minimizing complexity in the mode of
evolution \cite{13*}. Herrera and his collaborators \cite{15*}
formulated three complexity factors for an axially symmetric system
and examined a possible relation between symmetry and complexity of
the setup. They also employed this notion to establish a hierarchy
from the simplest (Minkowski) to more complex (radiating) systems
\cite{15a}. Sharif and Butt computed the complexity factor based on
Herrera's approach for a static cylindrically symmetric
self-gravitating system \cite{14a}. They also investigated the
effect of electromagnetic field on the complexity factor of static
spherical \cite{14*} as well as cylindrical \cite{14b} structures
and concluded that complexity increases in the presence of charge.
The complexity of a charged non-static spherical system has also
been explored \cite{14c}. Recently, Herrera et al. determined the
conditions under which a quasi-homologous system is complexity free
\cite{15b}.

Hubble's discovery of an expanding cosmos has been reconfirmed
through recent cosmological observations (redshift and
distance-luminosity relationship of type IA Supernovae \cite{14}).
Cosmological models in GR that explain the evolution of the universe
from its origin to its current phase suffer from some drawbacks
(like fine-tuning and cosmic coincidence problems). In order to find
an adequate solution, researchers modify GR to incorporate the
effects of accelerated expansion. Brans and Dicke \cite{15} modified
the Einstein-Hilbert action and developed a scalar-tensor theory
based on Dirac hypothesis and Mach principle. In Brans-Dicke (BD)
theory, the gravitational constant is replaced by a dynamical scalar
field ($\psi(t)=\frac{1}{G(t)}$) whose effect on matter is gauged
through a tunable coupling parameter $(\omega_{BD})$. The effect of
scalar field reduces corresponding to large values of the coupling
parameter.

The inflation of the universe is explained by lower values of
$\omega_{BD}$ \cite{17} whereas BD gravity is consistent in the
weak-field regime for $\omega_{BD}\geq40,000$ \cite{16}. In order to
establish a standard domain of the parameter, a potential function
($V(\Phi)$) is introduced. This function adjusts the values by
assigning a mass to the scalar field ($\Phi$) which leads to an
extension of BD gravity known as self-interacting BD (SBD) theory.
Sharif and Manzoor formulated structure scalars to study the
evolution of dynamical spheres \cite{22} and cylinders \cite{23'} in
SBD theory. Recently, the complexity of different geometries has
also been explored by employing Herrera's definition and it was
shown that complexity of the self-gravitating structures increases
in the presence of a massive scalar field \cite{100, 100a}. The
concept of complexity has been analyzed in other modified theories
as well \cite{23*}

In this paper, we derive the complexity factor for a dynamical
dissipative sphere by considering its pattern of evolution in the
background of SBD theory. The paper is organized as follows. In
section \textbf{2}, the SBD field equations and physical variables
representing a non-static sphere are evaluated. Structure scalars
are derived from the orthogonal splitting of the Riemann tensor in
section \textbf{3}. Section \textbf{4} gives an overview of the
complexity and evolution of the system. Kinematical quantities and
solutions corresponding to non-dissipative and dissipative fluids
are determined in section \textbf{5}. We discuss stability of the
vanishing complexity condition in section \textbf{6}. In the last
section, we summarize the main results.

\section{Self-interacting Brans-Dicke Theory and Matter Variables}

Self-interacting BD theory is defined via the action (with $8\pi
G_0=1$)
\begin{equation}\label{0}
S=\int\sqrt{-g}(\mathcal{R}\Phi-\frac{\omega_{BD}}{\Phi}\nabla^{\mu}\nabla_{\mu}\Phi
-V(\Phi)+\emph{L}_m)d^{4}x,
\end{equation}
where the Ricci scalar, determinant of metric tensor and matter
Lagrangian are represented by $\mathcal{R},~g$ and $\emph{L}_m$,
respectively. The SBD field and wave equations derived through the
variation of above action are, respectively, given as
\begin{eqnarray}\label{1}
G_{\mu\nu}&=&T^{\text{(\text{eff})}}_{\mu\nu}=\frac{1}{\Phi}(T_{\mu\nu}^{(m)}
+T_{\mu\nu}^\Phi),\\\label{2}
\Box\Phi&=&\frac{T^{(m)}}{3+2\omega_{BD}}+\frac{1}{3+2\omega_{BD}}
(\Phi\frac{dV(\Phi)}{d\Phi}-2V(\Phi)),
\end{eqnarray}
where the matter distribution is described by the energy-momentum
tensor $T_{\mu\nu}^{(m)}$ with $T^{(m)}=g_{\mu\nu}T_{\mu\nu}^{(m)}$.
The effects of massive scalar field are introduced in the matter
source through the following energy-momentum tensor
\begin{equation}\label{3}
T_{\mu\nu}^\Phi=\Phi_{,\mu;\nu}-g_{\mu\nu}\Box\Phi+\frac{\omega_{BD}}{\Phi}
(\Phi_{,\mu}\Phi_{,\nu}
-\frac{g_{\mu\nu}\Phi_{,\alpha}\Phi^{,\alpha}}{2})-\frac{V(\Phi)g_{\mu\nu}}{2},
\end{equation}
where $\Box\Phi=\Phi^{,\mu}_{~;\mu}$. We consider a collapsing
sphere bounded by a hypersurface $\Sigma$ represented in comoving
co-ordinates as
\begin{equation}\label{4}
ds^2=-A(r,t)^2dt^2+B(r,t)^2dr^2+R(t,r)^2(d\theta^2+\sin^2\theta
d\phi^2).
\end{equation}
The energy density ($\rho$), radial ($p_r$)/transverse ($p_\perp$)
pressures and heat flux $(q_\mu)$ of the anisotropic collapsing
sphere are specified by the following energy-momentum tensor
\begin{equation}\nonumber
T_{\mu\nu}^{(m)}=(\rho+p_\perp) u_{\mu}u_{\nu}+p_\perp
g_{\mu\nu}+(p_r-p_\perp)s_\mu s_\nu+q_\mu u_\nu+u_\mu q_\nu,
\end{equation}
where the 4-velocity ($u_\mu=(-A,0,0,0)$), radial 4-vector
($s_\mu=(0,B,0,0)$) and heat flux ($q_\mu=(0,qB,0,0)$) obey the
following relations
\begin{eqnarray*}
u^\mu u_\mu=-1,\quad u^\mu q_\mu=0,\quad s^\mu s_\mu=1,\quad s^\mu
u_\mu=0.
\end{eqnarray*}

In order to simplify the calculations, we introduce the quantities
\begin{eqnarray*}
\Pi_{\mu\nu}&=&\Pi(s_{\mu}s_{\nu}-\frac{h_{\mu\nu}}{3}), \quad
P=\frac{1}{3}(p_{r}+2p_{\perp}),\\
\Pi&=&p_r-p_\perp,\quad h_{\mu\nu}=g_{\mu\nu}+u_{\mu}u_{\nu},
\end{eqnarray*}
and rewrite the energy-momentum tensor as
\begin{equation}\label{5a}
T_{\mu\nu}^{(m)}=\rho
u_{\mu}u_{\nu}+Ph_{\mu\nu}+\Pi_{\mu\nu}+q(s_\mu u_\nu+u_\mu s_\nu).
\end{equation}
Using Eqs.(\ref{1})-(\ref{5a}), the field equations are obtained as
\begin{eqnarray}\label{6}
\frac{1}{\Phi}(A^2\rho-T_{00}^\Phi)&=&\frac{\dot{R}
\left(\frac{2\dot{B}}{B}+\frac{\dot{R}}{R}\right)}{R}-\frac{A^2
\left(\frac{R'^2}{R^2}-\frac{2B'R'}{BR}-\frac{B^2}{R^2}
+\frac{2R''}{R}\right)}{B^2},\\\label{7}
\frac{1}{\Phi}(-qAB+T_{01}^\Phi)&=&-\frac{2 A'\dot{R}}{AR}+\frac{2
\dot{B}R'}{BR}-\frac{2\dot{R}'}{R},\\\label{8}
\frac{1}{\Phi}(B^2p_r+T_{11}^\Phi)&=&-\frac{B^2\left(\frac{2
\ddot{R}}{R}-\frac{\dot{R}\left(\frac{2
\dot{A}}{A}-\frac{\dot{R}}{R}\right)}{R}\right)}{A^2}+\frac{R'
\left(\frac{2A'}{A}+\frac{R'}{R}\right)}{R}-\frac{B^2}{R^2},\\\nonumber
\frac{1}{\Phi}(R^2p_r+T_{22}^\Phi)&=&-\frac{R^2
\left(-\frac{\dot{A}\left(\frac{\dot{B}}{B}
+\frac{\dot{R}}{R}\right)}{A}+\frac{\dot{B}\dot{R}}{BR}
+\frac{\ddot{B}}{B(t,r)}+\frac{\ddot{R}}{R}\right)}{A^2}\\\label{9}
&+&\frac{R^2 \left(\frac{R'
\left(\frac{A'}{A}-\frac{B'}{B}\right)}{R}-\frac{A'B'}{A
B}+\frac{A''}{A}+\frac{R''}{R}\right)}{B^2},
\end{eqnarray}
where
\begin{eqnarray*}
T_{00}^\Phi&=&-\dot{\Phi}\left(\frac{2\dot{A}}{A}+\frac{\dot{B}}{B}+\frac{2
\dot{R}}{R}\right)+\frac{A^2\Phi'\left(\frac{B'}{B}+\frac{2
R'}{R}\right)}{B^2}+\frac{\omega_{BD}\left(\frac{A^2\Phi'^2}{B^2}
+\dot\Phi^2\right)}{2\Phi}\\
&+&\frac{A^2\Phi''}{B^2}+\frac{1}{2}V(\Phi)A^2,\\
T_{01}^\Phi&=&-\frac{A'\dot{\Phi}}{A}-\frac{\dot{B}\Phi'}{B}+\frac{\omega_{BD}}
{\Phi}\dot{\Phi}{\Phi}'+\dot{\Phi}',\\
T_{11}^\Phi&=&-\Phi'\left(\frac{A'}{A}+\frac{2B'}{B}+\frac{2
R'}{R}\right)+\frac{B^2\dot{\Phi}\left(\frac{\dot{A}}{A}+\frac{2
\dot{R}}{R}\right)}{A^2}+\frac{\omega_{BD}\left(\frac{B^2\dot{\Phi}^2}{A^2}
+\Phi'^2\right)}{2\Phi}\\
&+&\frac{B^2\ddot{\Phi}}{A^2}-\frac{1}{2} V(\Phi)B^2,\\\nonumber
T_{22}^\Phi&=&-\frac{R^2\Phi'\left(\frac{A'}{A}+\frac{B'}{B}+\frac{R'}{R}\right)}
{B^2}+\frac{R^2\dot{\Phi}\left(\frac{\dot{A}}{A}+\frac{\dot{B}}{B}
+\frac{\dot{R}}{R}\right)}{A^2}-\frac{\omega_{BD}R^2
\left(\frac{\Phi'^2}{B^2}-\frac{\dot{\Phi}^2}{A^2}\right)}{2\Phi}\\
&+&\frac{R^2
\ddot{\Phi}}{A^2}-\frac{R^2\Phi''}{B^2}-\frac{1}{2}V(\Phi)R^2.
\end{eqnarray*}
Here $'$ and $^{.}$ denote derivatives with respect to the radial
and temporal co-ordinates, respectively. The conservation equations
corresponding to the anisotropic matter source are expressed as
\begin{eqnarray}\nonumber
&&\dot{T}_0^{0(\text{eff})}+(T_0^{0(\text{eff})}-T_1^{1(\text{eff})})\frac{\dot{B}}{B}
+2(T_0^{0(\text{eff})}-T_2^{2(\text{eff})})\frac{\dot{R}}{R}+(T_0^{1(\text{eff})})'
\\\label{100}
&&+(T_0^{1(\text{eff})})(\frac{A'}{A}+\frac{B'}{B}+2\frac{R'}{R})=0,\\\nonumber
&&\dot{T}_0^{1(\text{eff})}+(T_1^{1(\text{eff})})'+T_0^{1(\text{eff})}(\frac{\dot{A}}{A}
+\frac{\dot{B}}{B}+2\frac{\dot{R}}{R})-(T_0^{0(\text{eff})}-T_1^{1(\text{eff})})\frac{A'}{A}
\\\label{101}
&&+2(T_1^{1(\text{eff})}-T_2^{2(\text{eff})})\frac{R'}{R}=0,
\end{eqnarray}
whereas the wave equation takes the following form
\begin{eqnarray}\nonumber
\Box\Phi&=&\frac{\Phi'\left(\frac{A'}{A}-\frac{B'}{B}+\frac{2
R'}{R)}\right)}{B^2}-\frac{\dot{\Phi}\left(\frac{-\dot{A}}{A}
+\frac{\dot{B}}{B}+\frac{2\dot{R}}{R}\right)}{A^2}
-\frac{\ddot{\Phi}}{A^2}+\frac{\Phi''}{B^2}\\\label{2*}
&=&\frac{1}{3+2\omega_{BD}}\left[-\rho+3P+
\left(\Phi\frac{dV(\Phi)}{d\Phi}-2V(\Phi)\right)\right].
\end{eqnarray}

Kinematical quantities (such as 4-acceleration $(a_\mu)$, expansion
scalar $(\Theta)$ and shear tensor $(\sigma_{\mu\nu})$) are used to
study the motion of cosmic objects. These quantities are defined as
\begin{equation}\nonumber
a_\mu=u_{\mu;\nu}u^{\nu}, \quad \Theta=u^\mu_{;\mu},\quad
\sigma_{\mu\nu}=u_{\mu;\nu}+a_{(\mu}u_{\nu)}-\frac{1}{3}\Theta
h_{\mu\nu},
\end{equation}
which for the considered setup turn out to be
\begin{eqnarray}\label{52}
a_1&=&\frac{A'}{A},\quad a^2=a_\mu
a^\mu=(\frac{A'}{AB})^2,\\\label{53}
\Theta&=&\frac{1}{A}(\frac{\dot{B}}{B}+2\frac{\dot{R}}{R}),\\\label{55}
\sigma_{11}&=&\frac{2}{3}B^2\sigma,\quad
\sigma_{22}=-\frac{1}{3}R^2\sigma,
\end{eqnarray}
with $a_\mu=as_\mu$ and
$\sigma=\sqrt{\frac{3}{2}\sigma^{\mu\nu}\sigma_{\mu\nu}}=\frac{1}{A}
(\frac{\dot{B}}{B}-\frac{\dot{R}}{R})$.

The boundary ($\Sigma$) of the fluid distribution divides the
spacetime into internal and external regions. In order to avoid a
discontinuity at the junction, the Darmois conditions must be
fulfilled. For this purpose, we assume that outgoing radiations are
massless as depicted in Vaidya spacetime given by
\begin{equation*}
ds^2=-(1-\frac{2M(\upsilon)}{r})d\upsilon^2-2rdrd\upsilon
+r^2(d\theta^2+\sin^2\theta d\phi^2),
\end{equation*}
where $M(\upsilon)$ and $\upsilon$ are the total mass and retarded
time, respectively.  The matching of the two spacetimes is smooth
and continuous when $(m(t,r))_{\Sigma}=(M(\upsilon))_{\Sigma},~
(q)_{\Sigma}=(p_{r})_{\Sigma},~(\Phi_-)_\Sigma=(\Phi_+)_\Sigma,~
(\Phi'_-)_\Sigma=(\Phi'_+)_\Sigma$ and
$(\dot\Phi_-)_\Sigma=(\dot\Phi_+)_\Sigma$ \cite{17a}. We use Misner
and Sharp \cite{18*} formula for calculating mass of the collapsing
model as
\begin{equation}\label{9a}
m=\frac{R^3}{2}\mathcal{R}^3_{232}=\frac{R}{2}\left[\left(\frac{\dot{R}}{A}\right)^2-
\left(\frac{{R'}}{A}\right)^2+1\right],
\end{equation}
where $\mathcal{R}^3_{232}$ is a component of Riemann tensor
$\mathcal{R}^{\alpha}_{\beta\gamma\delta}$. In order to discuss the
dynamics of the self-gravitating system, we introduce the proper
time and radial derivatives expressed as
\begin{equation*}
D_T=\frac{1}{A}\frac{\partial}{\partial t}, \quad
D_R=\frac{1}{R'}\frac{\partial}{\partial r}.
\end{equation*}
The velocity of the collapsing fluid in terms of aerial radius of
the spherical surface within the fluid is defined as $U=D_TR<0$. The
mass and velocity of the sphere are related as
\begin{equation}\label{57}
E\equiv\frac{R'}{B}=\left(1+U^2-\frac{2m}{R}\right)^\frac{1}{2}.
\end{equation}
Taking proper time and radial derivative of mass leads to
\begin{eqnarray}\label{59}
D_Tm=-\frac{R^2}{2}\left(\frac{T_{11}^{(\text{eff})}}{B^2}U
-\frac{T_{01}^{(\text{eff})}}{AB}E\right),\\\label{60}
D_Rm=-\frac{R^2}{2}\left(\frac{T_{00}^{(\text{eff})}}{A^2}
+\frac{T_{01}^{(\text{eff})}}{AB}\frac{U}{E}\right),
\end{eqnarray}
which imply
\begin{equation}\label{61}
\frac{3m}{R^3}=-\frac{T_0^{0(\text{eff})}}{2}+\frac{1}{2R^3}\int_0^r
R'R^3(D_RT_0^{0(\text{eff})}-
\frac{3T_{01}^{(\text{eff})}}{ABR}\frac{U}{E})dr.
\end{equation}

Tidal forces play a significant role in the evolution of a celestial
system. The Weyl tensor ($C^{\mu}_{\alpha\beta\sigma}$) incorporates
the effects of these forces and is expressed as
\begin{equation}\label{1'}
C^{\mu}_{\alpha\beta\sigma}=\mathcal{R}^{\mu}_{\alpha\beta\sigma}-\frac{\mathcal{R}^{\mu}_{\beta}}
{2}g_{\alpha\sigma}+\frac{\mathcal{R}_{\alpha\beta}}{2}\delta^{\mu}_{\sigma}
-\frac{\mathcal{R}_{\alpha\sigma}}{2}\delta^{\mu}_{\beta}
+\frac{\mathcal{R}^{\mu}_{\sigma}}{2}g_{\alpha\beta}+\frac{1}{6}(\delta^{\mu}_{\beta}
g_{\alpha\sigma}+g_{\alpha\beta}\delta^{\mu}_{\sigma}),
\end{equation}
where $~\mathcal{R}_{\alpha\beta}$ is the Ricci tensor. The Weyl
tensor is generally split into electric ($E_{\alpha\beta}$) and
magnetic ($H_{\alpha\beta}$) parts through the 4-velocity of the
observer. The magnetic part vanishes in spherical spacetime whereas
the electric part reads
\begin{equation}\label{11}
E_{\mu\nu}=C_{\mu\gamma\nu\delta}u^{\gamma}u^{\delta}=
\varepsilon(s_{\mu}s_{\nu}+\frac{h_{\mu\nu}}{3}),
\end{equation}
where
\begin{eqnarray}\nonumber
\varepsilon&=&\frac{1}{2}
\left(\frac{\left(\frac{R'}{R}-\frac{A'}{A}\right)
\left(\frac{B'}{B}+\frac{R'}{R}\right)+\frac{A''}{A}-\frac{R''}{R}}{B^2}
+\frac{\frac{\ddot{R}}{R}-\frac{\ddot{B}}{B}
-\left(\frac{\dot{A}}{A}+\frac{\dot{R}}{R}\right)
\left(\frac{\dot{R}}{R}-\frac{\dot{B}}{B}\right)}{A^2}\right.\\\label{14}
&-&\left.\frac{1}{R^2}\right).
\end{eqnarray}
Moreover, the relation
\begin{equation}\label{15}
[\varepsilon-\frac{1}{2}(-T_0^{0(\text{eff})}-T_1^{1(\text{eff})}+T_2^{2(\text{eff})})]^{\large{.}}
=\frac{3\dot{R}}{R}[\frac{1}{2}(-T_0^{0(\text{eff})}+T_2^{2(\text{eff})})-\varepsilon]-\frac{3R'}{2R}T_0^{1(\text{eff})}),
\end{equation}
demonstrates the influence of scalar field on energy density,
pressure and Weyl tensor.

\section{Structure Scalars}

We measure the complexity of the system through structure scalars
which are acquired from the orthogonal splitting of Riemann tensors.
These quantities were first evaluated by Herrera \cite{23}.
Following the same technique, we introduce the following tenors
\begin{eqnarray*}\label{23}
Y_{\alpha\beta}&=&\mathcal{R}_{\alpha\delta\beta\gamma}u^{\delta}u^{\gamma},\\\label{24}
Z_{\alpha\beta}=^*\mathcal{R}_{\alpha\delta\beta\gamma}u^{\delta}u^{\gamma}&=&
\frac{1}{2}\eta_{\alpha\delta\mu\epsilon}\mathcal{R}^{\mu\epsilon}_{\beta\gamma}
u^{\delta}u^{\gamma},\\\label{25}
X_{\alpha\beta}=^{*}\mathcal{R}^{*}_{\alpha\delta\beta\gamma}u^{\delta}u^{\gamma}&=&
\frac{1}{2}\eta_{\alpha\delta}^{\mu\epsilon}R^{*}_{\mu\epsilon\beta\gamma}
u^{\delta}u^{\gamma},
\end{eqnarray*}
where
$\mathcal{R}^{*}_{\alpha\beta\delta\gamma}=\frac{1}{2}\eta_{\mu\epsilon\delta\gamma}
\mathcal{R}_{\alpha\beta}^{\mu\epsilon}$ and
$^{*}\mathcal{R}_{\alpha\beta\delta\gamma}=\frac{1}{2}\eta_{\alpha\beta\mu\epsilon}
\mathcal{R}_{\delta\gamma}^{\mu\epsilon}$ are the right and left
duals, respectively. Using Eq.(\ref{1'}), the Riemann tensor can be
rewritten in the following form
\begin{equation}\label{26}
\mathcal{R}^{\alpha\delta}_{\beta\gamma}=C^{\alpha\delta}_{\beta\gamma}+
2T^{(\text{eff})[\alpha}_{[\beta}\delta^{\delta]}_{\gamma]}+
T^{(\text{eff})}
\left(\frac{1}{3}\delta^{\alpha}_{[\beta}\delta^{\delta}_{\gamma]}-
\delta^{[\alpha}_{[\beta}\delta^{\delta]}_{\gamma]}\right),
\end{equation}
and is decomposed as
\begin{equation}\label{27}
\mathcal{R}^{\alpha\delta}_{\beta\gamma}=\mathcal{R}^{\alpha\delta}_{(I)\beta\gamma}
+\mathcal{R}^{\alpha\delta}_{(II)\beta\gamma}+\mathcal{R}^{\alpha\delta}_{(III)\beta\gamma}
+\mathcal{R}^{\alpha\delta}_{(IV)\beta\gamma}+\mathcal{R}^{\alpha\delta}_{(V)\beta\gamma},
\end{equation}
where
\begin{eqnarray}\nonumber
\mathcal{R}^{\alpha\delta}_{(I)\beta\gamma}&=&\frac{2}{\Phi}\left[\rho
u^{[\alpha}u_{[\beta}\delta_{\gamma]}^{\delta]}
-Ph^{[\alpha}_{[\beta}\delta^{\delta]}_{\gamma]}+(\rho-3P)(\frac{1}{3}
\delta^{\alpha}_{[\beta}\delta^{\delta}_{\gamma]}-
\delta^{[\alpha}_{[\beta}\delta^{\delta]}_{\gamma]})\right],\\\nonumber
\mathcal{R}^{\alpha\delta}_{(II)\beta\gamma}&=&\frac{2}{\Phi}\left[
\Pi^{[\alpha}_{[\beta}\delta^{\delta]}_{\gamma]}
+q\left(u^{[\alpha}s_{[\beta}\delta^{\delta]}_{\gamma]}
+s^{[\alpha}u_{[\beta}\delta^{\delta]}_{\gamma]}\right)\right],\\\nonumber
\mathcal{R}^{\alpha\delta}_{(III)\beta\gamma}&=&4u^{[\alpha}u_{[\beta}E^{\delta]}_{\gamma]}
-\epsilon^{\alpha\delta}_{\mu}\epsilon_{\beta\gamma\nu}E^{\mu\nu},\\\nonumber
\mathcal{R}^{\alpha\delta}_{(IV)\beta\gamma}&=&\frac{2}{\Phi}\left[\Phi^{[,\alpha}_{[;\beta}
\delta^{\delta]}_{\gamma]}+\frac{\omega_{BD}}{\Phi}\Phi^{[,\alpha}\Phi_{[,\beta}
\delta^{\delta]}_{\gamma]}-\left(\Box\Phi+\frac{\omega_{BD}}{2\Phi}\Phi_{,\mu}
\Phi^{,\mu}+\frac{V(\Phi)}{2}\right)\right.\\\nonumber
&\times&\left.\delta^{[\alpha}_{[\beta}\delta^{\delta]}_{\gamma]}\right],
\\\nonumber
\mathcal{R}^{\alpha\delta}_{(V)\beta\gamma}&=&\frac{1}{\Phi}\left[\left(-\frac{\omega_{BD}}
{\Phi}\Phi_{,\mu}\Phi^{,\mu}-2V(\Phi)-3\Box\Phi\right)\left(\frac{1}{3}
\delta^{\alpha}_{[\beta}\delta^{\delta}_{\gamma]}-
\delta^{[\alpha}_{[\beta}\delta^{\delta]}_{\gamma]}\right)\right].
\end{eqnarray}
Here, we evaluate only $X_{\alpha\beta}$ and $Y_{\alpha\beta}$ as
\begin{eqnarray}\nonumber
X_{\alpha\beta}&=&\frac{1}{\Phi}\left(\frac{\rho
h_{\alpha\beta}}{3}+\frac{\Pi_{\alpha\beta}}{2}\right)-E_{\alpha\beta}
+\frac{1}{2\Phi}(\Phi_{,\alpha;\mu}h^{\mu}_{\beta}+
\frac{\omega_{BD}}{2\Phi}\Phi_{\alpha}\Phi_{\mu}h^{\mu}_{\beta})\\\label{32}
&+&\frac{h_{\alpha\beta}}{4\Phi}(\Box\Phi+7V(\Phi)),\\\nonumber
Y_{\alpha\beta}&=&\frac{1}{\Phi}\left(\frac{(\rho+3P)h_{\alpha\beta}}{6}+
\frac{\Pi_{\alpha\beta}}{2}\right)+E_{\alpha\beta}+\frac{1}{2\Phi}(-\Phi_{,\alpha;\beta}
-\Phi_{,\alpha;\mu}u_{\beta}u^{\mu}\\\nonumber
&-&\Phi_{,\mu;\beta}u_{\alpha}u^{\mu}
+\Phi_{,\gamma;\mu}u_{\gamma}u^{\mu}g_{\alpha\beta})
+\frac{\omega_{BD}}{2\Phi^2}(-\Phi_{,\alpha}\Phi_{,\beta}
-\Phi_{,\alpha}\Phi_{,\mu}u^\mu u_\alpha\\\nonumber
&-&\Phi_{,\mu}\Phi_{,\beta}u^\mu
u_\alpha-\Phi_{,\gamma}\Phi_{,\mu}u^{\gamma}u^\mu
g_{\alpha\beta})+\frac{h_{\alpha\beta}}{6\Phi}
\left(\frac{\omega_{BD}}{\Phi}\Phi_{,\mu}\Phi^{,\mu}-V(\Phi)
\right).\\\label{33}
\end{eqnarray}

The structure scalars appear in the trace and trace-free parts of
the above quantities as
\begin{eqnarray*}
X_{\alpha\beta}&=&\frac{X_T}{3}h_{\alpha\beta}+X_{<\alpha\beta>},\\
Y_{\alpha\beta}&=&\frac{Y_T}{3}h_{\alpha\beta}+Y_{<\alpha\beta>},
\end{eqnarray*}
where
\begin{eqnarray*}
X_T&=&X^{\alpha}_{\alpha},\quad
X_{<\alpha\beta>}=h^{\mu}_{\alpha}h^{\nu}_{\beta}\left(X_{\mu\nu}
-\frac{X^{\alpha}_{\alpha}}{3}h_{\mu\nu}\right),\\
Y_T&=&Y^{\alpha}_{\alpha},\quad
Y_{<\alpha\beta>}=h^{\mu}_{\alpha}h^{\nu}_{\beta}\left(Y_{\mu\nu}
-\frac{Y^{\alpha}_{\alpha}}{3}h_{\mu\nu}\right).
\end{eqnarray*}
The four structure scalars in the presence of scalar field turn out
to be
\begin{eqnarray}\nonumber
X_{T}&=&X_{T}^{\text{(m)}}+X_{T}^{\Phi}=\frac{1}{\Phi}(\rho)+\frac{1}{2\Phi}\left(\frac{5}{2}\Box\Phi
+\Phi_{,\alpha;\mu}u^{\alpha}u^{\mu}
+\frac{\omega_{BD}}{2\Phi}(\Phi_{,\alpha}\Phi^{,\alpha}\right.\\\label{35}
&+&\left.\Phi_{,\mu}\Phi_{,\alpha}u^{\alpha}u^{\mu}+\frac{21}{2}V(\Phi)\right),\\\nonumber
X_{TF}&=&X_{TF}^{\text{(m)}}+X_{TF}^{\Phi}=-\frac{1}{\Phi}(\frac{\Pi}{2}+\varepsilon\Phi)+\frac{1}{2\Phi}
\left(\Box\Phi+\Phi_{,\alpha;\mu}u^{\alpha}u^{\mu}+\frac{\omega_{BD}}
{2\Phi}(\Phi_{,\alpha}\Phi^{,\alpha}\right.\\\label{36}
&+&\left.\Phi_{,\mu}\Phi_{,\alpha}u^{\alpha}u^{\mu})\right),\\\nonumber
Y_{T}&=&Y_{T}^{\text{(m)}}+Y_{T}^{\Phi}=\frac{1}{2\Phi}(\rho+3p_r-2\Pi)
-\frac{1}{2\Phi}\left(\Box\Phi
+\Phi_{,\gamma;\alpha}u^{\gamma}u^{\alpha}\right.\\\label{37}
&+&\left.\frac{\omega_{BD}}{\Phi}(\Phi_{,\gamma}\Phi_{,\alpha}
u^{\gamma}u^{\alpha})+V(\Phi)\right),\\\nonumber
Y_{TF}&=&Y_{TF}^{\text{(m)}}+Y_{TF}^{\Phi}=\frac{1}{\Phi}(\varepsilon\Phi-\frac{\Pi}{2})
-\frac{1}{2\Phi} \left(\Box\Phi+\frac{\omega_{BD}}
{\Phi}(\Phi_{,\alpha}\Phi^{,\alpha}\right.\\\label{38}
&+&\left.\Phi_{,\gamma}\Phi_{,\beta}u^\gamma
u^\beta)+\Phi_{,\gamma;\mu}u^{\gamma}u^{\mu}\right).
\end{eqnarray}
The above equations indicate that $X_T$ and $Y_T$ govern the total
energy density and principal stresses of the system, respectively in
the presence of the massive scalar field. Moreover, $X_{TF}$ and
$Y_{TF}$ together determine the local anisotropy of the fluid. The
impact of anisotropy and inhomogeneity on the evolution of the
sphere can be measured through $Y_{TF}$ as
\begin{eqnarray}\nonumber
Y_{TF}&=&T_2^{2(\text{eff})}-T_1^{1(\text{eff})}+\frac{1}{2R^3}\int_0^r
R'R^3(-D_R
T_0^{0(\text{eff})}+\frac{3T_{01}^{(\text{eff})}}{ABR}\frac{U}{E})dr\\\label{39}
&+&\left[\frac{\dot{\Phi}}{A^2}(\frac{2\dot{A}}{A}+\frac{3\dot{R}}{R})
-\frac{3\Phi'R'}{B^2R}\right].
\end{eqnarray}

\section{Complexity and Evolution of the System}

According to the definition devised in \cite{13*}, the complexity of
the fluid distribution depends on the number of physical factors
required to adequately describe its structure. Thus, a spherical
object with dust fluid in its interior is less complex as compared
to the spherical structure consisting of a perfect fluid. In
general, the complexity of a cosmic system depends on various
physical properties such as anisotropic pressure and inhomogeneous
density. In \cite{100}, $Y_{TF}$ was chosen as the complexity factor
of the static sphere because it incorporated the essential features
of the system and determined their effects on Tolman mass (or active
gravitational mass). Equation (\ref{39}) indicates that $Y_{TF}$
contains the contribution of the significant factors which induce
complexity in the current setup. Therefore, we proceed by assuming
that the scalar $Y_{TF}$ is the best fit for the complexity factor.
Moreover, heat dissipation is an additional factor contributing to
the complexity of the dynamical setup. Therefore, it is essential to
take into account the pattern of evolution of the system to
construct a satisfactory complexity factor.  Furthermore, in order
to minimize the complexity, we will consider the anisotropic fluid
evolving through the simplest mode of evolution. For this purpose,
we identify two patterns of evolution: homologous and homogeneous.

\subsection{The Homologous Evolution}

The collapse of a celestial body is homologous if the rate at which
matter is pulled to the core is the same throughout, i.e., the
velocity of the matter falling inward is directly proportional to
the radial distance. On the other hand, if density at the center
increases rapidly as compared to other regions, then the cosmic
object evolves in a non-homologous pattern. In this section, we
derive the condition for a homologous collapse. Heat flow can be
expressed in terms of shear and expansion scalars through
Eqs.(\ref{7}) and (\ref{57}) as
\begin{equation}\label{58}
\frac{1}{2E\Phi}\left(q-\frac{T_{01}^\Phi}{AB}\right)
=\frac{1}{3}D_R(\Theta-\sigma) -\frac{\sigma}{R},
\end{equation}
which yields
\begin{equation}\label{62}
D_R\left(\frac{U}{R}\right)=\frac{1}{2E\Phi}
\left(q-\frac{T_{01}^{(\text{eff})}}{AB}\right)+\frac{\sigma}{R}.
\end{equation}
Integration of the above equation leads to
\begin{equation}\label{63}
U=R\int_0^rR'\left[\frac{1}{2E\Phi}
\left(q-\frac{T_{01}^{(\text{eff})}}{AB}\right)+\frac{\sigma}{R}\right]dr+c(t)R,
\end{equation}
where $c(t)=\frac{U_{\Sigma}}{R_{\Sigma}}$ is an integration
function. If the fluid is non-dissipative and shear-free then the
integral in the above equation vanishes providing the necessary
condition of homologous evolution $U\sim R$ \cite{110a}. Thus, the
ratio of aerial radii of any two concentric circles must be
constant. It is evident from the homologous condition that $R$ is a
separable function of $t$ and $r$. The homologous condition
corresponding to the current setup is
\begin{equation}\label{64}
\frac{1}{2E\Phi}
\left(q-\frac{T_{01}^{\Phi}}{AB}\right)+\frac{\sigma}{R}=0.
\end{equation}

\subsection{The Homogeneous Expansion}

The evolution of a cosmic structure is homogeneous if the rate of
expansion or collapse is independent of $r$. In other words,
homogeneous expansion corresponds to $\Theta'=0$. Applying this
constraint along with Eq.(\ref{64}) to (\ref{58}) implies
\begin{equation*}
D_R \sigma=0,
\end{equation*}
which leads to $\sigma=0$ (due to the regularity conditions at the
core). Thus, Eq.(\ref{58}) yields
\begin{equation}\label{65}
q=\frac{T_{01}^\Phi}{AB},
\end{equation}
i.e., the fluid is dissipative. It must be noted that in GR, a
shear-free matter distribution evolving under the condition
$\Theta'=0$ must also be non-dissipative and consequently,
homologous.

\section{Kinematical Variables}

In this section, we analyze the behavior of different physical
quantities to choose the simplest pattern of evolution. Imposing the
homologous condition on Eq.(\ref{58}) produces
\begin{equation}\nonumber
(\Theta-\sigma)'=\left(\frac{3\dot{R}}{AR}\right)'=0\Rightarrow
A'=0.
\end{equation}
Thus, the homologous fluid is geodesic ($a=0$) in the current
scenario. This implies that homologous pattern can be considered as
the simplest mode of evolution. Without loss of generality, we take
$A=1$. Conversely, the geodesic condition produces
\begin{equation*}
(\Theta-\sigma)=\frac{3\dot{R}}{R}.
\end{equation*}
Successive derivatives with respect to $r$ close to the center imply
that the fluid is homologous \cite{13*}.

It must be noted that the counterpart of this structure in GR is
shear-free when $q=0$. However, in the presence of scalar field, the
non-dissipative as well as homologous fluid is geodesic but not
shear-free as
\begin{equation*}
\sigma=\frac{RT_{01}^{\Phi}}{2R'}.
\end{equation*}
If the non-dissipative fluid undergoes homogeneous expansion, then
Eq.(\ref{65}) implies $T_{01}^{\Phi}=0.$ Moreover, shear scalar is
evaluated from Eq.(\ref{58}) as
\begin{eqnarray*}
\sigma&=&\frac{3}{2R^3}\int_0^r\frac{R^3}{A}T_{01}^{\Phi}dr+\frac{g(t)}{R^3}=\frac{g(t)}{R^3},
\end{eqnarray*}
where $g(t)$ is an arbitrary function of integration. Since at the
center $R=0$ therefore, $g(t)$ must be zero. It follows that in the
non-dissipative case, homogeneous expansion implies homologous
evolution (since $T_{01}^{\Phi}=0\Rightarrow\sigma=0\Rightarrow
U\sim R$). Conversely, if $\sigma=\frac{RT_{01}^{\Phi}}{2R'}$ then
$\Theta'=(\frac{RT_{01}^{\Phi}}{2R'})'$. Thus, homologous evolution
implies homogeneous expansion only if $T_{01}^{\Phi}=0$. In the
subsequent sections, we obtain solutions satisfying the conditions
for vanishing complexity as well as homologous fluid. For this
purpose, we assume an exponential form of the scalar field as
$\Phi(t,r)=\Phi(t)=\Phi_0t^b$, where $b$ is a constant and $\Phi_0$
is the present day value of the scalar field.

\subsection{Case 1: $q=0$}

We first consider the non-dissipative case. It is worthwhile to
mention here that the homologous fluid for the chosen scalar field
satisfies $T_{01}^{\Phi}=0$. Hence, in the non-dissipative case,
there is a unique criterion for the simplest evolution (since
homologous evolution fulfils the conditions of homogeneous expansion
and vice versa). The homologous condition yields
\begin{equation}\label{66}
B(t,r)=g_1(r)R(t,r),
\end{equation}
where $g_1(r)$ is an arbitrary function of integration. Employing
the above relation in the condition of vanishing complexity and wave
equation generates the following expressions
\begin{eqnarray*}
&&V(\Phi)=\frac{\Phi_0t^{\beta-2}}{h(r)^3R^4}\left[2t^2R'\left(h(r)
R'+h'(r)\right)-2t^2h(r)RR''+h(r)^3R^3\right.\\
&&\times\left.\left(t\left(5\beta\dot{R}+4 t\ddot{R}\right)+\beta
(\beta (\omega_{BD}+2)-2) R\right)+2th(r)^3
R^2\dot{R}\left(t\dot{R}+\beta R\right)\right],\\
&&\frac{\Phi_0t^{\beta-1}}{\beta(2\omega_{BD}+3)h(r)R}
\left[2t^2h'(r)\left(-2\beta R'R^2+R
\left(R'\left(t\dot{R}+\beta\right)+t \dot{R}'\right)\right.\right.\\
&&-\left.\left.4tR' \dot{R}\right)+2t^2h(r)\left(R^2\left(\beta
R''-t \dot{R}''\right)+t \left(3
R''\dot{R}+2R'\dot{R}'\right)R\right.\right.\\
&&-\left.\left.4tR'^2\dot{R}\right)+h(r)^3
R^2\left(-4t^3\dot{R}^3-\beta t^2\left(11\dot{R}^2+2\right)R+tR^2
\left(\beta(6
\beta\omega_{BD}\right.\right.\right.\\
&&+7\beta-7)\dot{R}\left.\left.\left.+t\left(5\beta
\ddot{R}+4t\dddot
R\right)\right)+2(\beta-2)\beta(\beta\omega_{BD}+\beta-1)
R^3\right)\right]=0.
\end{eqnarray*}
A complete solution can be determined for a suitable choice of
$g_1(r)$.

\subsection{Case 2: $q\neq0$}

In the non-dissipative case, the homologous, zero complexity and
wave equations, respectively, read
\begin{eqnarray*}
&&B=g_2(r)\exp\left(\int_1^t\frac{\Phi_0 t^{\beta}R\dot{R}'-R'
\dot{R}}{\left(\Phi_0t^{\beta}-1\right)RR'}\,dt\right),\\
&&V(\Phi)=\frac{\Phi_0 t^{\beta -2}}{B^3R}\left[2t^2 B'R'+2t
\dot{B}B^2\left(t\dot{R}+\beta
R\right)-2t^2BR''+B^3\left(t\left(5\beta
\dot{R}\right.\right.\right.\\
&&+\left.\left.\left.4t\ddot{R}\right)+\beta(\beta(\omega_{BD}+2)-2)R\right)\right],\\
&&\frac{\Phi_0 t^{\beta -1}}{\beta(2\omega_{BD}+3)B
R}\left[-6t^3B'\dot{B}RR'+2tB^3\left(-t^2
\dot{B}\dot{R}^2+tR\left(\dot{B}\left(t\ddot{R}-\beta
\dot{R}\right)\right.\right.\right.\\
&&+\left.\left.\left.t\ddot{B}\dot{R}\right)+\beta(\beta\omega_{BD}+\beta
-1)\dot{B}R^2\right)-2t^2B^2\left(R\left(t\dot{B}^2\dot{R}-\beta
R''+t\dot{R}''\right)\right.\right.\\
&&+\left.\left.\beta\dot{B}^2R^2-\beta R'^2-tR''
\dot{R}\right)+2t^2B\left(R\left(B'\left(t\dot{R}'-\beta R'\right)+t
\left(2\dot{B}R''\right.\right.\right.\right.\\
&&+\left.\left.\left.\left.\dot{B}'R'\right)\right)-tB'R'\dot{R}\right)+B^4
\left(t^2\left(-\left(7\beta\dot{R}^2+4t\ddot{R}\dot{R}+2\beta
\right)\right)+tR\left(\beta\right.\right.\right.\\
&&\times\left.\left.\left.(4\beta\omega_{BD}+5\beta-5)\dot{R}+t
\left(5\beta\ddot{R}+4t \dddot{R}\right)\right)+2(\beta-2)\beta
(\beta\omega_{BD}+\beta\right.\right.\\
&&\left.\left.-1)R^2\right)\right]=0,
\end{eqnarray*}
where $g_2(r)$ is an integration function. The above system of
equations provide a solution corresponding to an appropriate form of
$g_2(r)$ for $\Phi(t,r)=\Phi(t)=\Phi_0t^b$.

\section{Stability of $Y_{TF}=0$ Condition}

In this section, we examine whether the state of zero complexity can
prevail throughout the evolution of homologous matter distribution
for $\Phi(t,r)=\Phi(t)=\Phi_0t^b$. The evolution of the complexity
factor is obtained through Eqs.(\ref{100}) and (\ref{15}) as
\begin{eqnarray}\label{102}
\dot{Y}_{TF}+\frac{\dot{\Pi}}{\Phi}+\frac{3\dot{R}}{R}Y_{TF}+(\rho+P_r)\frac{\sigma}{2\Phi}
+\frac{1}{2B\Phi}(q'-\frac{qR'}{R})+\frac{2\Pi
\dot{R}}{R\Phi}+S_1=0,
\end{eqnarray}
where the term $S_1$, containing the effects of scalar field, is
given as
\begin{eqnarray*}
S_1&=&\frac{(T_1^{1\Phi}-T_2^{2\Phi})^{.}}{2\Phi}-\frac{(T_0^{1\Phi})'}{\Phi}
-\frac{(T_0^{1\Phi})'}{2\Phi}(\frac{B'}{B}-\frac{R'}{R})-
\frac{(T_0^{0\Phi}-T_1^{1\Phi})\dot{B}}{2B\Phi}\\
&-&\frac{5(T_0^{0\Phi}-T_2^{2\Phi})\dot{R}}{2R\Phi}-\dot{Y}_{TF}^{\Phi}-Y_{TF}.
\end{eqnarray*}

In the non-dissipative scenario, we assume that
$q=\Pi=\sigma=Y_{TF}=0$ at $t=0$ which leads to the following forms
of Eq.(\ref{102}) and its derivative with respect to $t$
\begin{eqnarray}\label{103}
S_1&=&-(\dot{Y}_{TF}+\dot{\Pi}),\\\label{104}
\ddot{Y_{TF}}+\frac{\ddot{\Pi}}{\Phi}-\frac{\dot{\Pi}\dot{\Phi}}{\Phi^2}&=&3S_1
-\dot{S_1}+\frac{\dot{\Pi}\dot{R}}{R\Phi}.
\end{eqnarray}
Employing the above relations, the first and second $t$-derivatives
of Eq.(\ref{39}) can be written as
\begin{eqnarray*}
S_1+3\left(\frac{\dot{\Phi}\dot{R}}{R}\right)^{.}&=&\frac{\partial}{\partial
t}\left(\int_0^rR^3(T_0^{0\text{(eff)}})'dr\right),\\
3S_1-\dot{S}_1+\frac{\dot{\Pi}\dot{R}}{R\Phi}-3\left(\frac{\dot{\Phi}\dot{R}}{R}\right)^{..}
&=&\frac{\partial^2}{\partial
t^2}\left(\int_0^r-R^3(T_0^{0\text{(eff)}})'dr\right).
\end{eqnarray*}
We can proceed in the same manner and calculate the higher
derivatives of Eq.(\ref{39}). It is noted that the stability of
vanishing complexity depends on state determinants (pressure and
energy density) as well as the massive scalar field.  Thus,
anisotropy and inhomogeneity in pressure and energy density,
respectively induce complexity in the system. For the general case,
i.e., when $q\neq0$, it can be clearly deduced from Eq.(\ref{102})
that heat dissipation is an additional factor influencing the
$Y_{TF}=0$ condition.

\section{Summary}

Many researchers have explored the dynamics and structure of
self-gravitating objects to gain insight into the mechanism of the
cosmos. However, the interdependence of physical features (such as
energy density, pressure, luminosity, etc.) as well as continuous
evolution of astrophysical objects lead to a complicated yet
intriguing system. The purpose of this work is to formulate a
definition of complexity for non-static systems in the framework of
SBD theory. We have considered an anisotropic radiating sphere with
inhomogeneous energy density. In order to determine the complexity
of the celestial system, we have employed Bel's technique to split
Riemann tensor. The resulting elements have yielded scalars that
govern the structure of the self-gravitating system. In order to
incorporate the dynamical aspect of the non-static regime, we have
considered two possibilities for the simplest pattern of evolution:
homologous and homogeneous modes. Finally, we have applied the
condition of vanishing complexity on homologous distribution to
formulate possible solutions for dissipative as well as
non-dissipative models. The factors due to which the system can
depart from zero complexity during the process of evolution have
also been discussed.

The structure scalars evaluated in SBD theory include the massive
scalar field and potential function which imply that the scalar
field contributes to the complexity of the system. Thus, the SBD
spherical system is more complicated than its GR counterpart. The
structure scalar $Y_{TF}$ has been selected as an appropriate choice
for complexity factor based on the following reasons.
\begin{itemize}
\item It has already served as an adequate measure of complexity
in the static case \cite{100}, thereby ensuring that the current
definition of complexity is recovered in the static regime.
\item It includes the effects of anisotropy, inhomogeneous energy
density and dissipation.
\end{itemize}
Since the homologous condition has implied that the fluid is
geodesic (for both $q=0$ and $q\neq0$) therefore, a homologous
pattern of evolution has been chosen to minimize the complexity in
the evolution of the system. It is interesting to mention here that
the homologous condition includes the effects of the scalar field.
Thus, in the non-dissipative case, the complexity factor and
shear-tensor do not vanish in contrast to the GR analog \cite{13*}.
The use of homologous and vanishing complexity conditions for
dissipative as well as non-dissipative models have provided open
systems that can be closed by choosing suitable integration
functions. Furthermore, we have deduced that in SBD gravity the
stability of vanishing complexity condition depends on the scalar
field in addition to the matter variables (pressure, heat flux,
energy density). It is noteworthy to mention here that all the
results are recovered for GR \cite{13*} under the conditions
$\Phi=\text{constant}$ and $\omega_{BD}\rightarrow \infty$.


\vspace{0.25cm}

\begin{thebibliography}{40}

\bibitem{1} Kolmogorov, A.N.: Prob. Inform. Theory J. \textbf{1}(1965)3;
Grassberger, J.: Int. J. Theor. Phys. \textbf{125}(1986)907;
Anderson, P.W.: Phys. Today \textbf{7}(1991)9; Parisi, G.: Phys.
World \textbf{6}(1993)42.

\bibitem{5} Lopez-Ruiz, R., Mancini, H.L. and Calbet, X.: Phys.
Lett. A \textbf{209}(1995)321; Calbet, X. and Lopez-Ruiz, R.: Phys.
Rev. E \textbf{63}(2001)066116; Catalan, R.G., Garay, J. and
Lopez-Ruiz, R.: Phys. Rev. E \textbf{66}(2002)011102; Sa\~{n}udo, J.
and Lopez-Ruiz, R.: Phys. Lett. A \textbf{372}(2008)5283.

\bibitem{9} Sa\~{n}udo, J. and Pacheco, A.F.: Phys. Lett. A \textbf{373}(2009)807;
Chatzisavvas, K.Ch. et al.: Phys. Lett. A \textbf{373}(2009)3901; de
Souza, R.A., de Avellar, M.G.B. and Horvath, J.E.: arXiv: 1308.3519;
de Avellar, M.G.B. et al.: Phys. Lett. A \textbf{378}(2014)3481.

\bibitem{13} Herrera, L.: Phys. Rev. D \textbf{97}(2018)044010.

\bibitem{13*} Herrera, L., Di Prisco, A. and Ospino, J.: Phys. Rev.
D \textbf{98}(2018)104059.

\bibitem{15*} Herrera, L., Di Prisco, A. and Ospino, J.: Phys. Rev.
D \textbf{99}(2019)044049.

\bibitem{15a} Herrera, L., Di Prisco, A. and Carot, J.: Phys. Rev.
D \textbf{99}(2019)124028.

\bibitem{14a} Sharif, M. and Butt, I.I.: Eur. Phys. J. C
\textbf{78}(2018)850.

\bibitem{14*} Sharif, M. and Butt, I.I.: Eur. Phys. J. C
\textbf{78}(2018)688.

\bibitem{14b} Sharif, M. and Butt, I.I.: Chinese J. Phys. \textbf{61}(2019)238.

\bibitem{14c} Sharif, M. and Tariq, S.: Mod. Phys. Lett. A \textbf{35}(2020)28.

\bibitem{15b} Herrera, L., Di Prisco, A. and Ospino, J.: Eur. Phys. J. C \textbf{80}(2020)631.

\bibitem{14} Perlmutter, S. et al.: Nature \textbf{391}(1998)51; Riess, A.G. et
al.: Astron. J. \textbf{116}(1998)1009.

\bibitem{15} Brans, C. and Dicke, R.H.: Phys. Rev. \textbf{124}(1961)3.

\bibitem{17} Weinberg, E.J.: Phys. Rev. D \textbf{40}(1989)3950.

\bibitem{16} Will C.M.: Living Rev. Rel. \textbf{4}(2001)4.

\bibitem{22} Sharif, M. and Manzoor, R.: Gen. Relativ. Gravit. \textbf{47}(2015)98;
Phys. Rev. D \textbf{91}(2015)024018.

\bibitem{23'} Sharif, M. and Manzoor, R.:  Astrophys. Space Sci. \textbf{359}(2015)17;
Commun. Theor. Phys. \textbf{68}(2017)39.

\bibitem{100} Sharif, M. and Majid, A.: Chin. J. Phys. 61(2019)38.

\bibitem{100a} Sharif, M. and Majid, A.: Indian J. Phys. (2020);  Int. J. Geom. Methods Mod. Phys.
\textbf{16}(2019)1950174.

\bibitem{23*} Abbas, G. and Nazar, H.: Eur. Phys. J. C
\textbf{78}(2018)510; ibid. 957; Astrophys. Space Sci.
\textbf{364}(2019)11; Sharif, M., Majid, A. and Nasir, M.M.M.: Int.
J. Mod. Phys. A \textbf{34}(2019)32; Zubair, M. and Azmat, H.: Int.
J. Mod. Phys. D \textbf{29}(2020)2; Yousaf, Z., Bhatti, M.Z. and
Naseer, T.: Phys. Dark Universe \textbf{28}(2020)100535.

\bibitem{17a} Darmois, G.: M\'{e}morial
des sciences math\'{e}matiques \textbf{25}(1927)58.

\bibitem{18*} Misner, C.W. and Sharp, D.H.: Phys. Rev. \textbf{136}(1964)B571.

\bibitem{23} Herrera, L. et al.: Phys. Rev. D \textbf{79}(2009)064025.

\bibitem{110a} Schwarzschild, M.: \emph{Structure and Evolution of the Stars} (Dover, 1958);
Kippenhahn, R. and Weigert, A.: \emph{Stellar Structure and
Evolution} (Springer-Verlag, 1990); Hansen, C. and Kawaler, S.:
\emph{Stellar Interiors: Physical Principles, Structure and
Evolution} (Springer-Verlag, 1994).
\end{thebibliography}
\end{document}